\documentclass[prb,aps,twocolumn]{revtex4}
\usepackage{amsmath}
\usepackage{bm}
\usepackage{epsfig}
\usepackage{graphicx}
\usepackage{amsfonts}
\usepackage{amssymb}

\begin{document}

\author{M.M.~Glazov$^{1,2}$\footnote{glazov@coherent.ioffe.ru} and K.V.~Kavokin$^{1,3}$}
\affiliation{$^1$ A. F. Ioffe Physico-Technical Institute, Russian
Academy of Sciences, 194021 St. Petersburg, Russia\\ $^2$ LASMEA,
CNRS/Universit\'{e} Clermont-Ferrand II ``~Blaise-Pascal~'', 24,
av des Landais, 63177, Aubiere, France\\ $^3$ LNMO INSA-LPMC CNRS,
FRE 2686, 135 Avenue de Rangueil, F-31077 Toulouse Cedex 4,
France}
\title{Cavity polaritons: Classical behaviour of a quantum parametric oscillator}

\begin{abstract}
We address theoretically the optical parametric oscillator based
on semiconductor cavity exciton-polaritons under a pulsed
excitation. A ``hyperspin'' formalism is developed which allows,
in the case of large number of polaritons, to reduce quantum
dynamics of the parametric oscillator wavefunction to the
Liouville equation for the classical probability distribution.
Implications for the statistics of polariton ensembles are
analyzed.
\end{abstract}
\maketitle

\section{Introduction}

Semiconductor microcavities with embedded quantum wells exhibit a rich variety
of unusual light-matter coupling effects. In the strong coupling regime
excitons and photons are mixed into a new kind of quasiparticles, known as
cavity exciton-polaritons \cite{1}. These particles inherit sharp energy
dispersion of cavity photons and strong interaction nonlinearities of
excitons. The bosonic nature of cavity polaritons allows to observe a number
of coherent phenomena such as stimulated scattering of exciton-polaritons and
even, perhaps, their Bose-condensation \cite{2,3}.

A non-parabolic shape of the lower dispersion branch of
exciton-polaritons having a sharp mininum and flat wings allows
for the so called parametric process when a pair of polaritons
from the pump scatter into the signal and idler states with both
energy and momentum conserved \cite{3,4,5} (see Figure 1). The
parametric scattering process was studied in detail experimentally
under continuous-wave (cw) pumping in Refs. \onlinecite{6,7,8,9}
and under pulsed excitation in Refs. \onlinecite{10,11,12}.
Previous theoretical treatment was based on the Heisenberg
equations \cite{13,14,15}. The quantum state of the ``pumped''
polaritons has been described by a classical non-fluctuating field
which is valid only in the case of cw pumping. This approach
allowed to obtain a closed set of equations, describing parametric
amplification. However, effects of the parametric oscillations,
when due to scattering polaritons from signal and idler states
return to the pump state are lost in this description. Moreover,
solutions obtained in Refs. \onlinecite{13,14,15} become unstable
above the stimulation threshold. The semi-classical treatment
based on the approach analogous to the methods used for the
description of the four-wave mixing phenomena presented in Refs.
\onlinecite{16,17} gives no insight into the microscopic effects.
The general theory of the parametric oscillator, relaxing the
above approximations, is lacking in the literature, to the best of
our knowledge.

The aim of this paper is to develop a general quantum formalism allowing an
analytical treatment of the parametric oscillator under the pulsed pumping,
both below and above the stimulation threshold, provided the total number of
polaritons is large. We introduce the hyperspin pseudovector whose components
describe the pair correlations between pump, signal and idler states. Starting
from the Heisenberg equations for hyperspin components and treating them as
non-correlated quantities we obtain classically stable trajectories for the
hypespin dynamics. Then, within an approach close to that used earlier by one
of us to describe the dynamics of large total spins of magnetic polarons
\cite{18}, we analyze the Schroedinger equation in the hyperspin space, and
show that the number of relevant variables in the equation can be reduced, and
its quasi-classical solution can be found. It turns out that the dynamics of
the parametric oscillator can be described by the Liouville equation where the
squared wavefunction of the parametric oscillator plays the role of a
classical distribution function. We show that polaritons pass from the pump to
the signal state with some delay, and that the populations of the pump, signal
and idler states demonstrate dumped oscillations about their average values.
We also obtain values of the second-order coherence in the steady-state regime
and demonstrate that parametric oscillations of polaritons between signal,
pump and idler lead to disappearance of the initial coherence.

\section{Theory}

We consider a typical experimental situation when the polaritons are excited
in the lower branch of dispersion under a ``magic angle'' (see Fig. 1). The
process of the parametric scattring of two pump polaritons into the signal and
idler states can be described by the following Hamiltonian
\begin{equation}\label{eq1}
\mathcal H = \mathcal H_{0}+ \mathcal H_{int},
\end{equation}
where the first term $\mathcal H_{0}=E_{s}a_{s}^{\dag}a_{s}+E_{p}a_{p}^{\dag}a_{p}%
+E_{i}a_{i}^{\dag}a_{i}$ is the free propagation term, $E_{k}$ ($k=s,p,i)$ are
the energies of signal, pump and idler states, and $a_{k}^{\dag}$ and $a_{k}$
are bosonic creation and annihilation operators for each state. The second
term in (\ref{eq1}), describing the polariton-polariton interaction, reads
\cite{1,13,14,15}:
\begin{equation}\label{eq2}
\mathcal H_{int}=V(a_{p}a_{p}a_{s}^{\dag}a_{i}^{\dag}+a_{s}a_{i}a_{p}^{\dag}a_{p}%
^{\dag}), %
\end{equation}
where $V$ is the constant of polariton-polariton interaction, the first term
in brackets describes scattering from the pump into the signal and idler, and
the last one accounts for the reverse process.

\begin{figure}[h]
\centering
\includegraphics[width=0.4\textwidth]{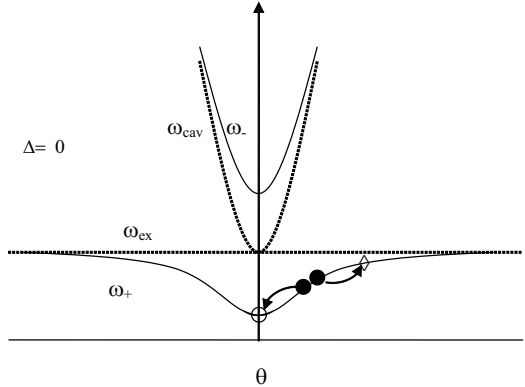}
\caption{Polariton energy ($\omega_{\pm}$) vs incident angle at
detuning $\Delta=0$ (solid). Cavity and exciton energies
($\omega_{cav}$, $\omega_{ex}$) are shown by dashed lines. Two
pump polaritons ({\textbullet}) scatter into signal ($o)$ and
idler ($\diamondsuit)$. (Adapted from Ref. \onlinecite{3}). }
\end{figure}

\subsection{Degenerate parametric oscillator}

To start with, we consider a simplified 2-level model, where
signal and idler states coincide [this model was invoked earlier to
describe experiments on degenerated 4-wave mixing \cite{dfwm}].
The interaction part of the Hamiltonian for this system can be
represented as
\begin{equation}\label{eq3}
\mathcal
H_{int}=V(a_{p}a_{p}a_{s}^{\dag}a_{s}^{\dag}+a_{s}a_{s}a_{p}^{\dag}a_{p}^{\dag
}). %
\end{equation}
It is well known that quantum mechanical description of two
selected states of one particle is possible in the terms of a
fictitious spin 1/2 (Ref. \onlinecite{Feynman}). This description
can be easily expanded to a system of $N$ bosons occupying two
quantum-mechanical states. Let us introduce operators
\begin{equation}\label{eq4}
X=\frac{1}{2}\left(  a_{s}a_{p}^{\dag}+a_{p}a_{s}^{\dag}\right)  %
\end{equation}%
\[
Y=-\frac{\mathrm i}{2}\left(
a_{s}a_{p}^{\dag}-a_{p}a_{s}^{\dag}\right)
\]%
\[
Z=\frac{1}{2}\left(  a_{p}^{\dag}a_{p}-a_{s}^{\dag}a_{s}\right)
\]
whose mean values give the difference of occupation numbers of the two states
($Z$) and second-order correlations between them ($X$ and $Y)$. These
operators obey the commutation relations of angular momentum components:
\begin{equation}\label{4a}
\lbrack X,Y]=\mathrm i Z,\quad\lbrack Z,X]=\mathrm i Y,\quad\lbrack Y,Z]=\mathrm i X %
\end{equation}

\begin{figure*}[t]
%
%
\centering
\includegraphics[width=0.9\textwidth]{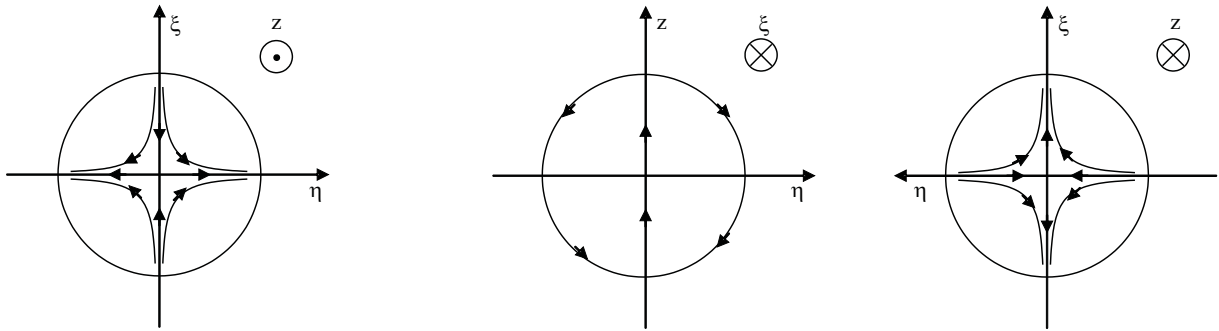}\newline \caption{Schematic
plots of the classically stable trajectories for the degenerate parametrical
oscillator. }%
\end{figure*}

Therefore, we introduce a pseudospin vector $\bm I$ with components $X$, $Y$,
$Z$ and the value $I=N$/2:
\begin{equation}\label{eq5}
X^{2}+Y^{2}+Z^{2} = I(I+1) = \frac{N^{2}}{4} + \frac{N}{2},
\end{equation}
where $N$ is the total number of polaritons.

For further treatment, it is more convenient to use another set of variables
$\xi,\eta,Z$ where $\xi$ and $\eta$ are defined as
\begin{equation}\label{eq6}
\xi=\frac{X+Y}{\sqrt{2}},\quad\eta=\frac{X-Y}{\sqrt{2}}. %
\end{equation}
The commutation relations for $\eta$ and $\xi$ are the same as for
$X$ and $Y$ respectively because $\eta$ and $\xi$ are obtained as
a result of rotation of the coordinate frame in the $(X,Y)$ plane.
In terms of new variables, the interaction part of the Hamiltonian
Eq. (\ref{eq3}) can be written as
\begin{equation}
\mathcal H_{int}=2V(\xi\eta+\eta\xi) \label{eq7}%
\end{equation}
Using commutation relations (Eqs. (\ref{4a}) and (\ref{eq6})) we obtain the
following Heisenberg equations for $\xi,\eta$ and $Z$ (dot means time
derivative)
\begin{equation}\label{eq8}
\dot{Z}=-\frac{4V}{\hbar}(\eta^{2}-\xi^{2}), %
\end{equation}%
\[
\dot{\eta}=\frac{2V}{\hbar}(Z\eta+\eta Z),
\]%
\[
\dot{\xi}=-\frac{2V}{\hbar}(Z\xi+\xi Z),
\]
Since the number of polaritons is very large, $I\gg 1$ and we can
treat the pseudospin as a classical vector and consider
trajectories of the end of this vector on a sphere with the radius
$I$ (see Fig. 2). Two meridional circles ($\eta=0$ and $\xi=0$)
are the stable trajectories, i.e. any trajectory starting near one
of the poles ($\eta=0,\xi=0,Z=I$ and $\eta=0,\xi=0,Z=-I$) very
soon comes very close to one of these circles. The equations of
motion on the meridional circles are exactly solvable. For
example, for $\xi=0$ we make a substitution $Z=I\cos{\varphi}$ and
$\eta=I\sin{\varphi}$ which gives a single equation for $\varphi$
\begin{equation}\label{eq9}
\dot{\varphi}=\frac{4VI}{\hbar}\sin{\varphi} %
\end{equation}
and its solution yields
\begin{equation}\label{eq10}
\varphi=2\arctan{\left[  \tan{\frac{\varphi_{0}}{2}}\exp{\left(
\frac
{4VI}{\hbar}t\right)  }\right]  } %
\end{equation}
where $\varphi_{0}$ is determined from the initial conditions. If the initial
condition corresponds to the correlated pump and signal states ($\eta$ is
non-zero), then $\varphi_{0}=\arcsin{(\eta/I)}\approx\eta/I$. If signal and
pump are initially not correlated, $\varphi_{0}=0$ and our approximation
evidently fails because all the time derivatives in Eqs. (\ref{eq8}) stay zero all
the time. In this important case we should go beyond the classical
approximation and take into account quantum fluctuations of pseudospin
components \cite{18}.

In order to do this we come back to the set of equations
(\ref{eq8}) but we will treat them as quantum equations for
operators $(\xi,\eta,Z)$. Using Eq. (\ref{eq5}) we can see that at
the initial moment, when $I-|Z|\ll I$, mean
square fluctuations of $\xi$ and $\eta$ are equal to $\langle\xi^{2}%
\rangle=\langle\eta^{2}\rangle=N/4+N/2(N/2-|Z|)$. At small $\xi$ and $\eta$,
$|Z|\approx I$ and is approximately constant. We can therefore simplify Eqs.
(\ref{eq8}) by putting $Z=I$, as follows
\begin{equation}\label{eq11}
\dot{\eta}=\frac{4V}{\hbar}I\eta, %
\end{equation}%
\[
\dot{\xi}=-\frac{4V}{\hbar}I\xi.
\]
These equations are exactly solvable and give an exponential
growth of $\eta$ and exponential decrease of $\xi$. So, at longer
time delays, when the approximate Eqs. (\ref{eq11}) are not valid,
we can still put $\xi\approx0$ and consider the motion along the
corresponding classically stable trajectory, but $\xi$, $\eta$ and
$Z$ should be treated as quantum-mechanical operators. The
commutation relations at $\xi\rightarrow0$ simplify:
\begin{equation}\label{eq12}
[\eta,Z] = 0, \quad [\xi, Z] = \mathrm i \eta, \quad [\eta, \xi] =
\mathrm i Z.
\end{equation}
Therefore, one can represent operators $Z$, $\eta$ and $\xi$ as
$\xi =-\mathrm i \partial/\partial\varphi$, $Z=\rho\cos{\varphi}$,
$\eta=\rho\sin{\varphi}$ where $\rho$ and $\varphi$ are classical
(not operatorial) variables. In this approximation, the
Hamiltonian (\ref{eq7}) corresponds to the following
time-dependent Schroedinger equation:
\begin{equation}\label{eq13}
\dot{\Psi}=-\frac{2V\rho}{\hbar}\left(
2\sin{\varphi}\frac{\partial\Psi
}{\partial\varphi}+\Psi\cos{\varphi}\right)  %
\end{equation}
This equation allows for explicit integration, namely
\begin{equation}\label{eq14}
\Psi^{2}\sin{\varphi}=F\left(  \frac{4V\rho}{\hbar}t-\ln{\left|
\tan
{\frac{\varphi}{2}}\right|  }\right)  , %
\end{equation}
where $F$ is a function to be determined from the initial conditions. It is
worth noting that $|F|d\varphi=|\Psi|^{2}dZ$ gives a probability distribution
over $Z$ for the degenerate parametric oscillator.

Eq. (\ref{eq14}) shows that in the limit of large number of polaritons, $N
\gg1$, the dynamics of our system is essentially classical and its
wavefunction can be replaced by the classical probability distribution, which,
in turn, satisfies the Liouville equation \cite{19}. Actually, comparing the
arguments of the function $F$ in Eq. (\ref{eq14}) and classical solutions for
$\varphi(t)$ given by Eq. (\ref{eq10}), one can see that the probability
distribution $F(\varphi,t)$ at any time moment $t$ can be obtained from the
initial one, $F_{0}(\varphi_{0})$, as $F(\varphi,t) = F_{0}(\varphi(t))$,
where $\varphi(t)$ is found by solving the classical system (\ref{eq8}).
In other words, one needs just to solve equations (\ref{eq8}), treating $Z$,
$\xi$, and $\eta$ as \textit{classical} variables, for a set of different
initial conditions with the statistics reflecting the initial $|\Psi|^{2}$,
and then the probabilities for all the pseudospin components at any time can
be obtained.

The validity of our approximation can be checked if we notice that
on a short time-scale after an initial moment our system is
equivalent to the harmonic oscillator with the inverted potential
($\mathcal H_{int}\propto p^{2}-x^{2}$, where $p$ and $x$ are
generalized momentum and coordinate respectively). For such a
problem the quasi-classical approximation is valid when the spread
of the wavefunction is larger than the oscillator length (root
mean square of $x$ in parabolic potential) \cite{19}. In our case,
when almost all the polaritons are initially in the pump state, the
wavefunction at $t=0$ is an eigenfunction of the operator $Z$. On
the other hand, this function is the first eigenfunction of the
harmonic oscillator. So, initially, the width of the wavefunction
is the same as the oscillator length, but in-plane components of
the pseudopsin rapidly increase with time (see Eq.(12)), and the wavefunction
spreads into the region of validity of the quasi-classical model.
This means that the dynamics of the pseudospin is quasi-classical
at any time except the short initial time range, where the system
obeys analytically solvable Eqs.(12).

The above analysis refers to the first half-period of the
hyperspin motion. For the second half-period the same arguments
can be used with the only difference that $\eta \to0$ serves as
the momentum operator and $\xi$ becomes the classical coordinate
variable.

\subsection{General case}

The description of the non-degenerate parametric oscillator (with different
signal and idler states) is analogous to the previous case. Here, to deal with
three polariton states, we introduce a 9-dimensional hyperspin pseudovector.
Its components describe correlations between all pairs of states. For example,
``signal-pump'' components are
\begin{equation}\label{eq15}
X_1 = \frac{1}{2}\left(a_s a_p^\dag +  a_p a_s^\dag\right)
\end{equation}
\[
Y_1 = -\frac{\mathrm i}{2}\left(a_s a_p^\dag -  a_p
a_s^\dag\right)
\]
\[
Z_1 = \frac{1}{2}\left(a_p^\dag a_p -  a_s^\dag a_s\right)
\]
The remaining components $X_{2},Y_{2},Z_{2}$ describing correlations between
pump and idler and $X_{3},Y_{3},Z_{3}$ for the idler-signal correlations can
be obtained by respective changing of indices. As it can be readily seen from
the definition, $Z_{1}+Z_{2}+Z_{3}=0$. The commutation relations for the
hyperspin components with the same index were already obtained in the previous
section. Other components obey the following relations
\begin{equation}\label{eq16}
[X_\alpha, X_{\alpha+1}] = [Y_{\alpha+1}, Y_\alpha] =
\frac{\mathrm i}{2} Y_{\alpha-1},
\end{equation}
\[
[X_\alpha, Y_{\alpha+1}] = [Y_\alpha, X_{\alpha+1}] =
\frac{\mathrm i}{2} X_{\alpha-1},
\]
\[
[X_\alpha, Z_{\alpha+1}] = [X_{\alpha}, Z_{\alpha-1}] =
\frac{\mathrm i}{2} Y_{\alpha},
\]
\[
[Y_\alpha, Z_{\alpha+1}] = [Y_{\alpha}, Z_{\alpha-1}] = -
\frac{\mathrm i}{2} X_{\alpha},
\]
\[
[Z_\alpha, Z_{\alpha+1}] = 0.
\]
where $\alpha \in \left\{ 1,2,3\right\} $, $\alpha \pm 3\equiv \alpha $.
The following combination plays the role of the squared total angular momentum:
\begin{equation}\label{eq17}
\sum_{i}(X_{i}^{2}+Y_{i}^{2}+\frac{2}{3}Z_{i}^{2})=\frac{N^{2}}{3}+N.
\end{equation}

The interaction Hamiltonian can be expressed in terms of the hyperspin
operators in a very simple form
\begin{equation}
\mathcal H_{int}=2V(X_{1}X_{2}+Y_{1}Y_{2}).
\end{equation}
Free propagation terms can be written as
\begin{equation}
\mathcal H_{0}=\frac{2}{3}\left[  (E_{p}-E_{i})Z_{1}+(E_{i}-E_{p})Z_{2}+(E_{s}%
-E_{i})Z_{3}+\right.
\end{equation}
\[
+\left.  (E_{s}+E_{p}+E_{i})\frac{N}{2}\right]  .
\]

Our next step is to obtain a set of Heisenberg equations for the hyperspin
operators. Using commutation relations (\ref{eq16}) and simple algebra we come
to
\begin{equation}\label{eq18}
\dot Z_{1} = - \frac{{3V}}{\hbar}(X_{1} Y_{2} - Y_{1} X_{2} ),\mathrm{{ }}%
\end{equation}%
\[
\displaylines{
\dot Z_2  = \frac{{3V}}{\hbar}(X_1 Y_2  - Y_1 X_2 ), \cr
\dot X_1  = \frac{V}{\hbar}(X_1 Y_3  + Y_1 X_3  + 2Z_1 Y_2 ) - \frac
{{E_p  - E_s }}{\hbar}Y_1 , \cr
\dot X_2  =  - \frac{V}{\hbar}(X_3 Y_2  + Y_3 X_2  - 2Y_1 Z_2 ) - \frac
{{E_i  - E_p }}{\hbar}Y_2 , \cr
\dot Y_1  =  - \frac{V}{\hbar}(Y_1 Y_3  - X_1 X_3  + 2Z_1 X_2 ) + \frac
{{E_p  - E_s }}{\hbar}X_1 , \cr
\dot Y_2  =  - \frac{V}{\hbar}(X_3 X_2  + Y_3 Y_2  + 2X_1 Z_2 ) + \frac
{{E_i  - E_p }}{\hbar}X_2 , \cr
\dot X_3  =  - \frac{V}{\hbar}%
(X_1 Y_1  + Y_1 X_1  - X_2 Y_2  - Y_2 X_2 ) + \frac{{E_i  - E_s }}{\hbar
}Y_3 , \cr
\dot Y_3  =  - \frac{V}{\hbar}(X_1^2  - Y_1^2  + Y_2^2  - X_2^2 ) - \frac
{{E_i  - E_s }}{\hbar}X_3 . \cr}
\]

Further treatment of this system is very similar to one we applied in the
degenerate case. The hyperspin system (\ref{eq18}) has a classically stable
trajectory defined by the following conditions:
\begin{equation}\label{eq18a}
Y_{2}=X_{1},\quad X_{2}=-Y_{1},\quad Y_{1}=\pm X_{1}. %
\end{equation}

The same argumentation as in the Sec. IIA shows that in the limit of large
number of polaritons the motion of system is concentrated near this
trajectory. This fact allows us to reduce number of relevant variables. In
order to do it, it is convenient to use another set of variables, namely
\begin{equation}\label{eq19}
\eta_{\pm}=\frac{1}{2}(X_{1}+Y_{2}\pm X_{2}\mp Y_{1}), %
\end{equation}%
\[
\xi_{\pm}=\frac{1}{2}(X_{2}+Y_{1}\pm X_{1}\mp Y_{2}),
\]%
\[
\zeta_{\pm}=\frac{1}{2}(Z_{1}-Z_{2}\mp X_{3}).
\]
On the classically stable trajectory defined by Eq.(\ref{eq18a}),
$\xi_{\pm }=0$, $\eta_{\pm}=X_{1}\mp Y_{1}$ and
$\zeta_{\pm}=Z_{1}\mp X_{3}/2$. These variables are remarkable by
the fact that the commutation relations of the variables with the
same index (+ or $-$) are the same as commutation rules for the
operators of the angular momentum:
\begin{equation}\label{eq20}
[\eta_\pm, \xi_\pm] = \mathrm i \zeta_{\pm}, \quad [\xi_\pm,
\zeta_\pm] = \mathrm i \eta_\pm, \quad [\eta_\pm, \zeta_\pm] =
-\mathrm i \xi_\pm,
\end{equation}
while the commutators of variables with different indices, and of
any of them with $Y_{3}$, are either exactly, or approximately
equal to zero in the vicinity of the classicaly stable trajectory
Eq. (\ref{eq18a}). When the hyperspin is close to the classically
stable trajectory, the commutator of $\eta_{\pm}$ and
$\zeta_{\pm}$ becomes negligible and the following substitution is
possible
\begin{equation}\label{defzeta}
\zeta_{\pm}=\rho_{\pm}\cos{\varphi_{\pm}},\quad\eta_{\pm}=\rho_{\pm}%
\sin{\varphi_{\pm}},
\end{equation}
where $\rho_{\pm}$ and $\varphi_{\pm}$ can be treated as $c$-numbers. To
satisfy first two relations in Eq. (\ref{eq20}), one should take
\begin{equation}
\xi_{\pm}=-\mathrm i \frac{\partial}{\partial\varphi_{\pm}}.
\end{equation}

In terms of the new variables, the interaction Hamiltonian Eq.(\ref{eq2}) is
decomposed in two parts describing independent degenerate parametric
oscillators
\begin{equation}
H_{int}=V\sum_{j=\pm}(\eta_{j}\xi_{j}+\xi_{j}\eta_{j})=- \mathrm i
V\sum_{j=\pm}\rho_j \left(
2\sin{\varphi_{j}}\frac{\partial}{\partial\varphi_{j}}+\cos{\varphi_{j}%
}\right)  .
\end{equation}
Here we assume that the parametric
oscillator wavefunction $\Psi$ depends on
$\rho_{\pm}$ and $\varphi_{\pm}$. At the exact resonance, $E_{p}-E_{s}%
=E_{i}-E_{p}$ and free propagation term in the Hamiltonian may be omitted,
since all possible configurations of the parametric oscillator have equal
energies. The solution of the corresponding Schoedinger equation is a function
of two arguments, namely
\begin{equation}\label{eq21}
\Psi^{2}\sin{\varphi_{+}}\sin{\varphi_{-}}= %
\end{equation}%
\[
F\left(  \frac{2V\rho_{+}}{\hbar}t-\ln{\left|  \tan{\frac{\varphi_{+}}{2}%
}\right|  },\frac{2V\rho_{-}}{\hbar}t-\ln{\left|  \tan{\frac{\varphi_{-}}{2}%
}\right|  }\right)  .
\]
The function $F$ is determined from initial conditions in the same way as for
the two-level model. The physical meaning of Eq. (\ref{eq21}) is exactly the
same as in the case of the two-level model: we can see that, for a large
number of polaritons, the wavefunction can be replaced by a classical
probability distribution which satisfies the Liouville theorem and evolves in
accordance to the set of dynamical equations for hyperspin components
(Eq.(\ref{eq18})).

\subsection{Finite polariton lifetimes}

The dynamical system considered above is idealysed as it does not
include dissipation processes. In reality,
polaritons have finite lifetimes dependent on the quality factor
of the cavity and on the non-radiative broadening of the exciton.
The decay of polaritons can be taken into account
phenomenologically within the hyperspin approach, by adding linear
dissipation terms to Eqs.(\ref{eq18}), in analogy to Bloch
equations. In the most general case, to keep linear dependence of
the decay rate on the population of polariton states, such terms
should have the following form: the total polariton population
(which has been so far assumed constant) is now given by the
equation $\dot{N}=-\alpha N-2\bm\beta\cdot\bm I$, where $\alpha$
and all the components of the vector $\bm\beta$ are constants; and
the dissipation of the hyperspin vector is determined by the term
$-\frac{1}{2}N\bm\beta-\alpha\bm I+\hat{\delta}\bm I$, where $\hat
{\delta}$ is a second-rank tensor. (We remind that $\bm I$ is the
9-dimensional hyperspin pseudovector.) The parameters $\alpha$,
$\bm\beta$, and $\hat{\delta}$ should be chosen in such a way as
to provide given values of the inverse lifetimes $\gamma_{k}$
($k=s,p,i$) of polaritons in signal (s), pump (p), and idler (i)
states for any direction of the hyperspin. This condition is met
if
\begin{equation}\label{lifetimes}
\alpha=1/3(\gamma_{s}+\gamma_{p}+\gamma_{i}),
\end{equation}
\[
\beta_{Z_{1}}=1/3(\gamma_{p}-\gamma_{s}),
\]
\[
\beta_{Z_{2}}=1/3(\gamma _{i}-\gamma_{p}),
\]
\[
\beta_{Z_{3}}=1/3(\gamma_{s}-\gamma_{i}),
\]
and the components of tensor $\hat \delta$ are
\begin{equation}
\delta
_{Z_{1},Z_{2}}=-\delta_{Z_{3},Z_{2}}=\beta_{Z_{2}},\quad \delta_{Z_{1},Z_{3}%
}=-\delta_{Z_{2},Z_{3}}=-\beta_{Z_{3}},
\end{equation}
\[\delta_{Z_{2},Z_{1}}=-\delta _{Z_{3},Z_{1}}=-\beta_{Z_{1}},\]
and all the other components of $\bm\beta$ and $\hat{\delta}$ are
zero.

\section{NUMERICAL RESULTS AND DISCUSSION}
\begin{figure}[h]
%
%
\centering
\vspace{20mm}\includegraphics[width=0.4\textwidth]{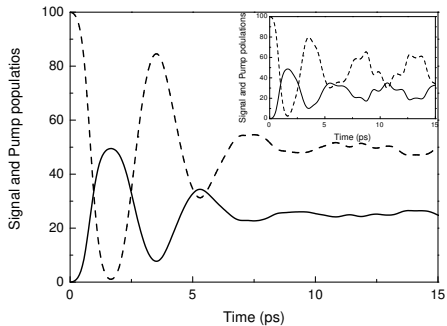}\newline\vspace{-30mm}
\caption{Signal (solid) and pump (dashed) populations as a
function of time. Idler population coincides with signal one. Main
figure corresponds to the results obtained in the hyperspin
formalism, and inset shows populations obtained by the direct
diagonalization of the interaction hamiltonian. The difference in
the behavior at time large than 7 ps may be attributed to the low
($N=100$) number of
particles in our simulations. }%
\end{figure}

\begin{figure*}[t]
%
%
\centering
\vspace{5mm}\hspace{10mm}\includegraphics[width=0.9\textwidth]{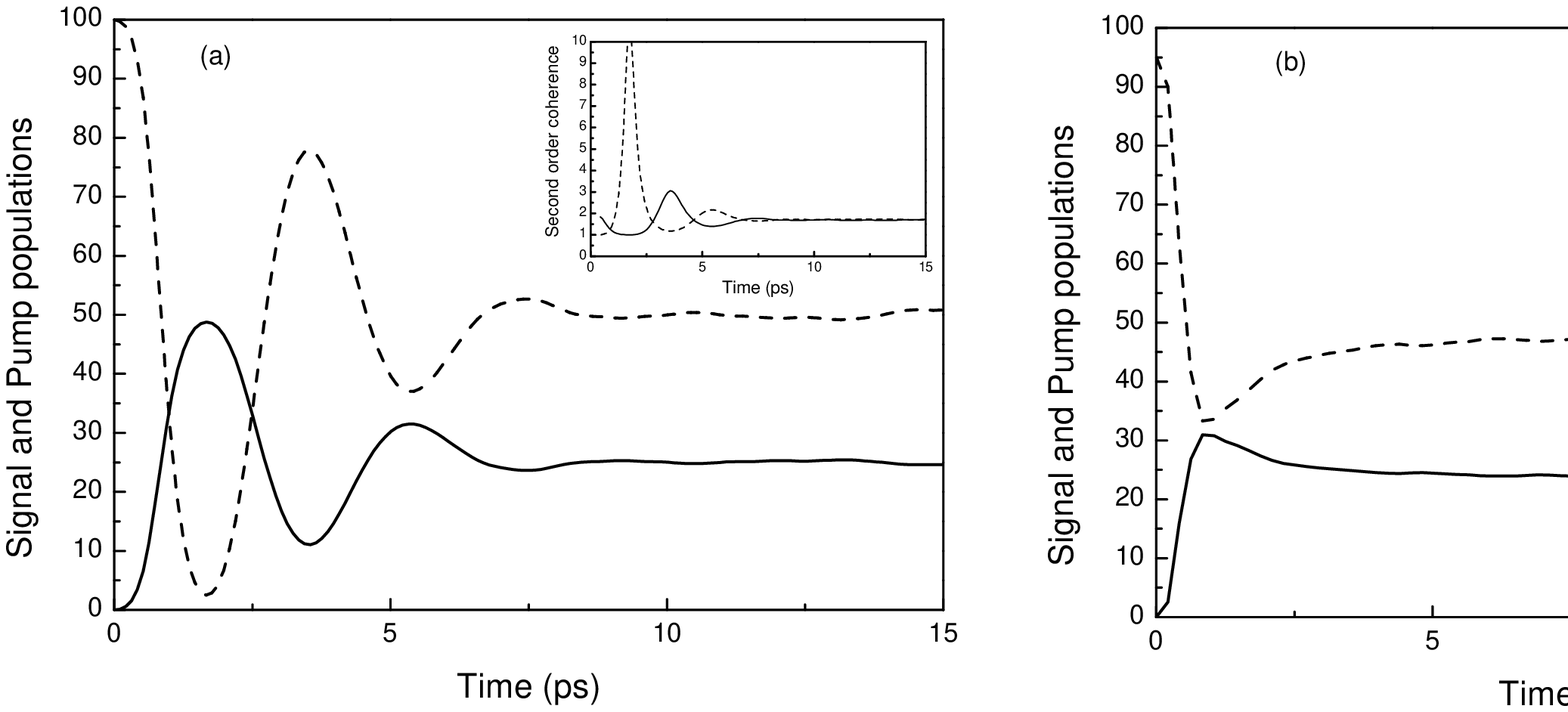}\newline\vspace{-65mm}
\caption{Time dependence of the signal (solid) and pump (dashed)
populations in the case of coherent (a) and thermal (b) pumping
without lifetime. Inset shows second-order coherence for the
signal state by solid line and for the pump
state by dashed line. }%
\end{figure*}

We will focus on the situation when the pump state was initially populated
while signal and idler states are empty. First of all, we will consider the pump
state with a definite number of particles (Fock state). The probability
distribution for $X_{i}$ and $Y_{i}$ components is given by
\begin{equation}\label{init}
\left|  \Psi(t=0)\right|  ^{2} =
\end{equation}
\[
\frac{1}{(2\pi)^{3}\sigma_{1}^{2} \sigma_{3}} \exp{\left(  - \frac{X_{1}%
^{2}+X_{2}^{2}+Y_{1}^{2}+Y_{2}^{2}}{2\sigma_{1}} - \frac{X_{3}^{2}+Y_{3}^{2}%
}{2\sigma_{3}} \right)  },
\]
where $\sigma_{1}$ and $\sigma_{3}$ are the mean square values of the
respective hyperspin components. The $Z$-projections of the hyperspin take
their maximum values $Z_{1,2} = \pm N/2$, resulting in $\sigma_{3}=0$ and
$\sigma_{1} = N/4$ (see Eq. (\ref{eq17})).

Figure 3 shows the dynamics of the signal and pump populations
calculated within our hyperspin formalism compared to those
obtained by direct diagonalization of the Hamiltonian
Eq.(\ref{eq2}) (see inset). The fully numerical solution of the
Schoedinger equation with the Hamiltonian given by Eq. (\ref{eq2})
has been carried out by projecting the Hamiltonian operator to the
basis of the Fock states with the definite number of particles in
signal, pump and idler. The eigenvalues and eigenvectors of the
Hamiltonian matrix were found numerically making possible to
construct a time-dependent solution in the form of a Fourier
expansion. Using this time-dependent wavefunction it is possible
to calculate all relevant values, namely, the average populations
of states and their second order coherence (see below). Within
this approach, the total number of polaritons, $N$, cannot be
taken very large. We performed calculations with $N=100$, which
provided reasonable computation time and, on the other hand,
allowed the comparison with our quasiclassical hyperspin model
valid at large $N$. Calculations within the pseudospin approach
were performed by solving Eqs. (\ref{eq18}) for a set of initial
conditions distributed randomly according to Eq. (\ref{init}),
with a subsequent averaging.  We took interaction constant
$V=2.1\times 10^{-2}$~meV so as to make the product $VN\sim2$~meV
close to the realistic value \cite{1}. Populations of signal and
pump states demonstrate damped oscillations centered at $N/4$ and $N/2$
respectively. Figure 3 clearly
demonstrates two main features of parametric oscillations: (a) the
pump polaritons pass to signal and idler states and vice versa,
and (b) polaritons arrive to the signal state with some delay with
respect to the initial moment when the pump state was excited, as
it was shown experimentally in Ref. \onlinecite{10}.

The quarter-period of the oscillations $T_{1/4}$ can be found
analytically from the solution of the Schroedinger equation
obtained in a previous section. For simplicity we consider an
evolution of variables $\varphi_+$ and $\zeta_+$ only, assuming
that the initial distributions in $+$ and $-$ variable sets are
the same. According to Eq. (\ref{defzeta}), the quarter of the
period corresponds to $\varphi_+ = \pi/2$. On the other hand, if
the initial value of $\varphi_+$ is $\varphi_0$, then
\begin{equation}
\frac{2V\rho_+}{\hbar} t - \ln{\left|  \tan{\frac
{\varphi_+(t)}{2}}\right|  } = -\ln{\left|  \tan{\frac
{\varphi_0}{2}}\right|  }.
\end{equation}
(This relation can be derived comparing the argument of the
function $F$ in Eq. (\ref{eq21}) at two time moments $0$ and $t$).
Thus the quarter of period at fixed $\varphi_0$ is given by
\begin{equation}\label{quarter}
t_{1/4} = \frac{\hbar}{2V\rho_+} \ln{\left|  \tan{\frac
{\varphi_0}{2}}\right|  }.
\end{equation}
Eq. (\ref{quarter}) should be averaged with the initial
distribution of $\varphi_0$. Reminding that $\varphi_0 \ll 1$,
one can put $\varphi_0 \approx \eta_0 /\rho_+$ where $\eta_0$ is
the initial value of $\eta_+$ and $\rho_+ = N/2$. The variable
$\eta_+ = X_1 - Y_1$ has initially Gaussian distribution with the
standard deviation $\sigma_\eta = 2\sigma_1 = N/2$. Finally,
\begin{equation}
T_{1/4}=\int_{-\infty}^{\infty}\frac{1}{\sqrt{2\pi\sigma_{\eta}%
}}\exp{\left(  -\frac{\eta^{2}}{2\sigma_{\eta}}\right)
}t_{1/4}=%
\end{equation}
\[
=\frac{\hbar}{VN}\left[  \ln{\left(
\frac{N}{\sqrt{2\sigma_{\eta}}}\right) }+\frac{\gamma}{2}\right]
\approx0.9\mbox{ }ps
\]
where $\gamma\approx0.577216$ is the Euler constant. This value
concides with the result of numerical calculation
$T_{1/4}\approx0.9$~ps (see Fig. 3).

\begin{figure*}[t]
%
%
\centering
\includegraphics[width=0.9\textwidth]{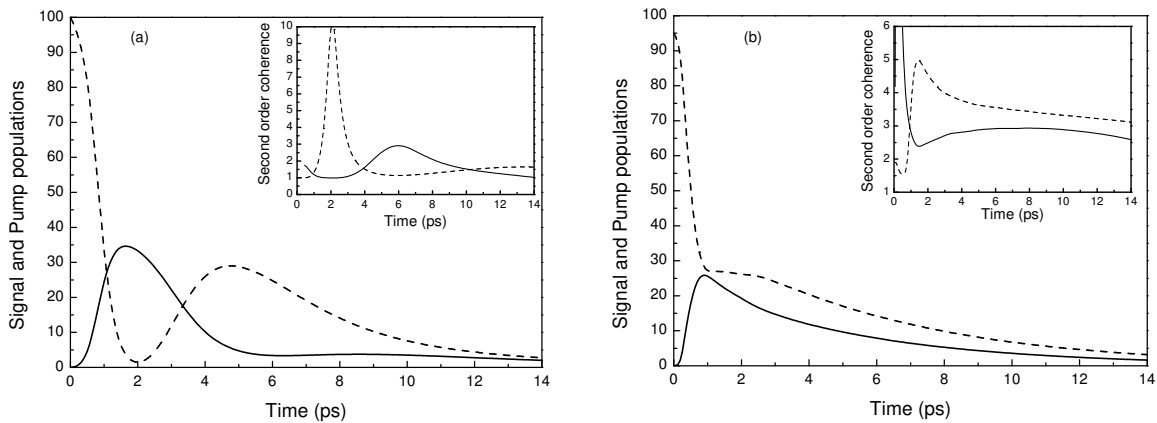}\newline\vspace{-60mm} \caption{Time
dependence of the signal (solid) and pump (dashed) populations in
the case of coherent (a) and thermal (b) pumping with lifetime
$5$~ps. Inset shows second-order coherence for the signal state by
solid line and for the pump
state by dashed line. }%
\end{figure*}

The damping of the oscillations results from the fact that the
Hamiltonian (\ref{eq2}) has almost continuous spectrum at $N\gg1$.
The summation of oscillations on different eigen-frequencies leads
to the degradation and damping of the oscillations of the signal,
idler and pump populations. The steady state values can be
obtained from the detailed balance, i.e. the numbers of incoming
and outcoming particles for each state should be equal. Since
polaritons can arrive to or leave the pump state by pairs only
(see Fig. 1a) and both signal and idler were not populated
initially, then the steady state values for the pump, $N_{p}$, and
signal, $N_{s}$, states obey the equations $N_{p}+2N_{s}=N$ and
$2N_{s}=N_{p}$ with the solutions $N_{p}=N/2$ and $N_{s}=N/4$.

One can see that at time $t<7$~ps the hyperspin formalism and direct
diagonalization give utterly similar results. The discrepancies at longer time
delays are due to the fact that the number of particles ($N=100$) we used for
simulation was too small for the quasi-classical approximation to hold on the
larger time scale.

Now we proceed to discussion of quantum statistics of the
parametric oscillator. We consider two different cases of initial
pumping: coherent with
initial pump statistics $P(N_{p}) = \alpha^{2N_{p}} \exp{(-\alpha^{2})}%
/N_{p}!$, where $\alpha^{2}$ is the average number of particles
and thermal one with initial statistics $P(N_{p}) =
(1-\theta)\theta^{N_{p}}$, where $\theta=
\bar{N}_{p}/(1+\bar{N}_{p})$ and $\bar N_{p}$ is the average
number of polaritons in the pump state. Figure 4 presents time
dependence of the signal and pump populations and so-called
second-orded coherence $g_{k}^{2} = \langle a_{k}^{\dag2}
a_{k}^{2}\rangle/\langle a_{k}^{\dag} a_{k}\rangle^{2}$ (with
$k=s,p$) which can be measured in the two-photons counting
experiments. Its value characterizes the statistics of the
relevant state and equals to $1$ for the coherent state and to $2$
for the thermal one. It can be seen that in the case of coherent
initial pump the time dependences of the populations are similar
to those obtained for the pump in the Fock state (see Fig. 4(a)
and Fig. 3 for comparison). The case of thermal pumping (Fig.
4(b)) is different. The oscillations are suppressed due to the
large spread of the initial distribution in the particle number
and the populations of the states reach their steady values almost
monotonously. Our calculations show that the steady state values
of $g_{s}^{2}$ and $g_{p}^{2}$ are very close in magnitude and are
$1.7$ in the case of initially coherent pump and $3.3$ for the
thermal pump. These values are confirmed by the direct
diagonalization of the hamiltonian ($N=100$) which gives $1.7$ and
$3.2$ respectively. Such values are different from those obtained
in Ref. \onlinecite{15} for constant wave pumping. It may seem
strange that coherence dissapears in the interacting system.
However this result becomes clear if one notices that in the case
of the polariton laser the spontaneous coherence build-up is
possible if the incoming scattering rate to the ground state is
larger than the outgoing scattering rate \cite{20}. In our case of
a three level system, incoming and outgoing rates are actually the
same and the coherence is destroyed by multiple scattering
processes.

The calculated populations and statistics for the polaritons with
finite lifetime are presented in Figure 5. The lifetimes for all
states were choosen identical and equal to $\gamma^{-1}=5$~ps. One
can see in the case of initially coherent pump (Fig. 5a) that the
populations of signal and pump states behave nonmonotonously. From
the very beginning polaritons start passing from the pump state to
the signal and idler states, while initially this process goes
spontaneously, and signal population growth slowly. Then,
stimulated scattering switches on, and both signal and pump
populations change rapidly. Then the process of return of
polaritons from signal and idler to the pump state starts, etc.
Because of the finite lifetime, the total number of particles in
the system decreases. When it falls below the stimulation
threshold $N\sim\hbar/V\gamma$ no more oscillations of the
occupation numbers of signal and pump states can be seen.

The situation is different in the case of the thermal pump (see Fig. 5b): no
pronounced oscillations can be seen for the pump population, however the sharp
bend of the time dependence of the pump population and the increase of the
signal population demonstrate that the parametric process takes place nevertheless.

Interestingly, introduction of lifetime in the system results in better
coherence at $t=14$~ps $g_{s}^{2}\approx1$ in the case of coherent pumping and
in almost the same value $g_{p}^{2}\approx3$ for the initially thermal pump.

In conclusion, we have presented the general formalism describing
the dynamics of the optical parametric oscillator based on a
semiconductor microcavity in the strong coupling regime in the
case of large number of polaritons. The introduction of the
hyperspin allows to obtain a quasi-classical solution of the
quantum parametric-oscillator problem. We have shown, that the
probability distribution for the hyperspin components obeys the
Liouville equation. Our approach is shown to give very good
agreement with the method based on the direct diagonalization of
the Hamiltonian for a ``moderate'' number of particles ($N=100$)
allowing such a numerical solution within a reasonable computation
time. At the same time, our quasiclassical hypespin model
radically reduces the computational complexity of the problem for
large number of particles, gives a possibility to introduce finite
lifetimes of polaritons, and allows to obtain some analytical
results such as an expression for the period of oscillations. The
presented numerical results suggest that the polariton parametric
oscillator can be used as a system to grow coherence in the case
where irreversible processes are present. Our formalism can be
extended to allow for spin degree of freedom of polaritons and
might therefore be suitable for description of polarization
properties of the microcavity emission in the parametric regime.

\begin{acknowledgments}
We thank A. Kavokin, I.
Shelykh, M.Dyakonov, T.Amand, and P.Renucci for valuable
discussions, and D. Solnyshkov for help in numerical computation.
This work was supported by the European Research Office of the US
Army, contract N62558-03-M-0803, by ACI Polariton project, and by
the Marie Curie MRTN-CT-2003-503677 ``Clermont 2''.
\end{acknowledgments}

\end{document}